\documentclass[aps,prb,twocolumn]{revtex4}
\usepackage{color}
\usepackage{graphicx}
\usepackage{amssymb}
\usepackage{epstopdf}
\def\be{\begin{equation}}
\def\ba{\begin{array}}
\def\bfg{\begin{figure}}
\def\ef{\end{figure}}
\def\bea{\begin{eqnarray}}		
\def\ee{\end{equation}}
\def\ea{\end{array}}
\def\eea{\end{eqnarray}}		
\def\nn{\nonumber \\}

\begin{document}

\title{Electron-hole spectra created by adsorption on metals from density-functional theory}
\author{Matthias Timmer}
\author{Peter Kratzer}
\affiliation{Fachbereich Physik - Theoretische Physik and Centre for Nanointegration (CeNIDE), Universit\"at
Duisburg-Essen, Lotharstr.~1, 47048 Duisburg, Germany}
\email{peter.kratzer@uni-due.de}
\date{\today}                                          
\begin{abstract}
Non-adiabaticity in adsorption on metal surfaces gives rise to a number of measurable effects, such as chemicurrents and exo-electron emission. 
Here we present a quantitative theory of chemicurrents on the basis of ground-state density-functional theory (DFT) calculations of the effective electronic potential and the Kohn-Sham band structure. 
Excitation probabilities are calculated both for electron-hole pairs and for electrons and holes separately from first-order time-dependent perturbation theory. This is accomplished by evaluating the matrix elements (between Kohn-Sham states) of the rate of change of the effective electronic potential between subsequent (static) DFT calculations. 
Our approach is related to the theory of electronic friction, but allows for direct access to the excitation spectra. 
The method is applied to adsorption of atomic hydrogen isotopes on the Al(111) surface. 
The results are compatible with the available experimental data (for noble metal surfaces); in particular, the observed isotope effect in H versus D adsorption is described by the present theory. 
Moreover, the results are in qualitative agreement with computationally elaborate calculations of the full dynamics within time-dependent density-functional theory, with the notable exception of effects due to the spin dynamics. 
Being a perturbational approach, the method proposed here is simple enough to be applied to a wide class of adsorbates and surfaces, while at the same time allowing us to extract system-specific information. 
\end{abstract}


\maketitle

\section{Introduction}
When an atom or molecule interacts with a metal surface, it generates electronic excitations in the metal. 
This general statement is true because the metal offers a continuum of electronic excitations, reaching down to arbitrary small excitation energies, and hence the adiabatic limit for scattering or adsorption of particles at a metal surface is ill-defined.\cite{Ande67}
Dissipation of some part of the particle's energy to the metal surface via electronic excitation has been known for a long time, and is the physical cause of such phenomena as vibrational damping of adsorbate vibrations on metals,\cite{HePe84} exo-electron emission in chemical reactions on metal surfaces,\cite{Greb97,Nien02} and the so-called chemicurrents (see below). 
Moreover, for the sticking of atoms or molecules to a surface, it is necessary that the adsorption energy is dissipated to the substrate. 
Thus the question arises what role is played by electronic excitation processes in the energy dissipation during sticking. 
The ubiquity of (albeit small) non-adiabatic contributions to chemical reactions at metal surfaces underlines the practical importance of understanding the process of electronic dissipation.\cite{Hass07} 
So far, most calculations addressing gas-surface interactions assume that an adiabatic regime would exist, by treating the interaction dynamics on a Born-Oppenheimer potential energy surface. 
While this assumption may often yield plausible results, it is strictly speaking not justified for metals, as some electronic excitations must always occur, as discussed above. 
On the basis of calculations for dissociative adsorption of hydrogen molecules, it has been argued recently that the energy dissipated to the electronic system is rather small,\cite{JuAl08} even on a metal, and thus has little effect e.g. on the energy dependence of the sticking coefficient.\cite{NiPi06} These findings seem to suggest that some (probably the larger) part of the adsorption energy of the sticking particles is dissipated to excitations of the lattice. 
However, further experimental clarification of this point is desirable. 
The measurement of chemicurrents offers an alternative possibility to quantify the amount of electronic excitation that takes place during chemisorption. In these experiments, the gas-phase particles are adsorbed on a metal film, fabricated either by previous deposition on a semiconductor substrate or a thin insulating barrier layer, forming a Schottky diode~\cite{Nien02} or a metal-insulator-metal (MIM) device,\cite{MiHa06} respectively. 
While particles are being adsorbed, an electric current is detected, originating from carriers that have been excited sufficiently 
strongly to overcome the in-built potential barrier of the detection device. Typical barrier heights are of the order 0.4 -- 1~eV. 
The physics of the underlying processes is similar to the emission of exo-electrons, only differing in the fact that the lower internal barrier, e.g. in a Schottky diode, allows for detection of a larger fraction of the excited carriers as compared to emission into the vacuum.
The latter process is possible solely for electrons whose energy exceed the work function of the metal, typically several eV. 
The size of the measured chemicurrents is 
dependent on the chemical nature of the adsorbing species, 
and to a lesser extent also on the metal surface.\cite{GeNi01}
The strong observed isotope effect (so far only data for hydrogen versus deuterium adsorption are available, differing by about a factor three)\cite{KrNu07b} clearly shows that the origin of the measured chemicurrents is related to {\em dynamical} electronic excitations.
The size of the isotope effect is, so far, the most sensitive test for a theoretical description of chemicurrents, 
as it reflects the small high-energy tail of the spectrum of electronic excitations. 

In this paper, we are going to present a theory of chemicurrents that allows us to make material-specific predictions about the spectrum of the electronic excitations and the isotope effect. To begin with, one has to keep the following observations in mind: 
While electronic excitations obviously contribute to the overall energy dissipation, previous calculations~\cite{TrGr02,LiPe06b,JuAl08} have shown that just {\em a small fraction} of the chemisorption energy is dissipated while the particle approaches the chemisorption potential well and makes the first round trip. Hence the influence of the electronic energy dissipation on the particle's trajectory is a small (second-order) effect.
Moreover, the vast majority of the electrons and holes produced by the excitation has an excitation energy too small to overcome the barrier of the Schottky diode and get detected.
Hence, the efficiency of an incident particle for creating detectable excited carriers in these experiments is tiny, less than one percent.\cite{GeNi01,KrNu07b} Therefore, we suggest 
the application of time-dependent perturbation theory to calculate these small excitation probabilities, using as input the realistic electronic structure of the metal substrate obtained from density-functional theory calculations.

Previously, there have been several other approaches in use to calculate chemicurrents: 
In recent work by Lindenblatt and Pehlke, the dynamics of the full electronic system of a slab model of the surface has been calculated from time-dependent density functional theory (TDDFT).\cite{LiPe06b} This approach is conceptually very appealing, but computationally cumbersome. Moreover, it requires high numerical accuracy, as the excitation energy of the electronic system, a small number, needs to be calculated by comparing huge numbers, the total internal energies in the ground state and the excited state. 
In a different approach, the energy dissipated to the electronic system has been calculated from the theory of electronic friction,\cite{TrGr01,TrGr02} that was originally developed in the context of adsorbate vibrational damping. However, this theory does not yield directly the spectra of the excited carriers. Instead, it is assumed that the excitation is made up from a superposition of individual electron-hole pairs of small excitation energies. These energies are assumed to be limited to a narrow energy region around the Fermi energy of the metal, allowing for a calculation of the excitation probability solely based on the properties of electronic states right at the Fermi level. For higher excitation energies, the probability is calculated by referring to the additional assumption of independent multiple electron-hole pair excitations that can be described by a Poissonian process.\cite{TrBi03} Thus, the assumptions made do not allow 
one to take into account the electronic single-particle properties of the 
metal under study further away from the Fermi energy.
Finally, yet another theory has been worked out that allows 
the calculation of chemicurrents by solving the dynamics of an appropriate model system, described by a Newns-Anderson Hamiltonian.\cite{MiBi05,MiBi07,MiBi07E,BiMi08,MiBi08} This scheme is computationally lighter than the full TDDFT dynamics, and allows 
the discussion of interesting trends over a variety of model systems.\cite{MiBi07,MiBi07E} Material-specific physics can be described by matching model parameters to values calculated from DFT;\cite{BiMi08} however, no detailed account of the metallic band structure for the excitation of chemicurrents is attempted in this theory.
 
The time-dependent perturbation theory worked out in this paper puts us in position to calculate electron and hole excitation spectra separately and directly from the Kohn-Sham band structure and wavefunctions obtained in DFT. From the spectra, the excitation efficiency per impinging particle and its dependence on isotopic mass can be calculated without further drastic assumptions. The method allows 
us to make material-specific predictions while at the same time being computationally affordable, as only data from ground-state DFT calculations are required. 
The remainder of this paper is organized as follows: First, an outline of the theory, including a discussion of its limitations, is given. Secondly, results are presented for a specific system, H/Al(111), where we can compare to the results of full TDDFT calculations from the literature. Finally, we discuss the relationship of our present theory with other approaches and conclude.

\section{Theory}
In this section we present the perturbative approach used for calculating the spectrum of electrons and holes excited by an adsorbing particle. 
The motion of the particle is described by its classical trajectory.
For simplicity, our notation assumes a one-dimensional motion along a reaction coordinate $Q(t)$, but generalisation to multiple dimensions is straightforward. 
Typically the particle's trajectory will make many oscillations in the adsorption well before it has dissipated the adsorption energy to the substrate. In the following,
we consider just the first round trip of the trajectory in the adsorption potential in the time interval between $-\tau$ and $\tau$. For conceptual clarity, we assume that a trajectory unaffected by energy dissipation can be used. 
This assumption is a good first approximation in many cases of practical interest, but it can be lifted without much difficulty if required.
For a full description of adsorption, one should decompose the oscillating trajectory into many round trips, and use a somewhat different energy and corresponding starting point for each segment of the trajectory. In our case the energy loss during the first round trip is already sufficient for the particle to become trapped.

According to the Runge-Gross theorem,\cite{RuGr84} time-dependent density functional theory (TDDFT) 
establishes a mapping of the electronic many-particle problem onto an effective single-particle Hamiltonian of the form
\bea
\label{Hamiltonian_Veff}
H^{\sigma}(t)  = - \frac{\hbar^2}{2m_{e}} \nabla^{2} + V_{\rm{eff}}^{\sigma}(t).
\eea
The effective potential $V_{\mathrm{eff}}(t)$ consists of 
\bea
\label{Veff}
V_{\mathrm{eff}}^{\sigma}(t) = V_H(t) + V_{XC}^{\sigma}(t) + V_{\mathrm{ion}}(t).
\eea
$V_{\mathrm{ion}}$ depends on time implicitly through the particle's position $Q(t)$.
The charge density needs to be calculated self-consistently from the 
solution of the time-dependent Kohn-Sham equations. 
Hence the Hartree and exchange-correlation potentials, $V_H$ and $V_{xc}$, are time-dependent, too.

While TDDFT is in principle exact, we are seeking for suitable approximations to arrive at a computationally simpler scheme. 
With this motivation, 
we assume that the true effective potential $V_{\mathrm{eff}}(t)$ in the time-dependent Kohn-Sham equations can be approximated by the effective potential $V_{\mathrm{eff}}(Q)$ of a time-independent problem, namely, the electronic ground state for a system with the particle located at $Q(t)$. In view of the density of excited electrons and holes being small compared to the overall charge density, this approximation appears plausible. However, there are cases where deviations from the instantaneous ground state are strong; in particular those system where exo-emission has been observed. From the theory point of view, a class of system can be envisaged where 
the adsorbate, in some points of its trajectory, must be described (at least in an approximative sense) by quantum numbers (such as charge state, spin, etc.) that differ from those of the instantaneous ground state.
For these systems, we anticipate that one may need to go beyond the perturbative approach outlined here.

The change of the effective potential due to the scattering of an incident particle constitutes a localized time-dependent perturbation.
The chemicurrent consists of propagating charge carriers that are detected (in principle) far away from the particle's impact point. 
These carriers are described by the eigenstates $| \varepsilon_i \rangle$ of the Hamiltonian $H_0 := H(Q_0)$. 
In order to describe vibrational damping, $Q_0$ should be taken as the adsorption height.  In the case of interest here, we take $Q_0 \to - \infty$, i.e., $H_0$
describes the (semi-infinite) metal with the particle far away from the surface.
This defines a decomposition of the Hamiltonian into an unperturbed part $H_0$ and a perturbation
$
V(Q) = V_{\mathrm{eff}}(Q) - V_{\mathrm{eff}}(-\infty) 
$
that depends on time implicitly through $Q(t)$.
While $V(Q)$ is generally {\em not} small, its  perturbative treatment relies on the fact that $V(Q(t))$ varies rapidly and is different from zero only within a short time interval.
Here, we are interested in electron-hole pair excitations above some threshold energy $h \nu_{S} \gtrsim 0.5$~eV due to the high-frequency components of the perturbation. For the validity of the perturbative treatment, it is required that the response of the electronic system on the time scale set by $\nu_{S}^{-1}$ is sufficiently weak.
Applying first-order time-dependent perturbation theory, the transition amplitude for an excitation of an electron from an occupied state $i$ into an unoccupied state $j$ is given by
\be
p_{ij}(t) =  \left\langle \varepsilon_{j} \left| V(Q(t)) \right| \varepsilon_{i} \right\rangle \exp(i(\varepsilon_{j}-\varepsilon_{i}) t / \hbar ),
\ee
and the excitation spectrum is obtained from
\be
P_{\mathrm{ex}} (\omega) = \sum_{ij} \left| \int_{-\tau}^{\tau} \frac{dt}{\hbar} p_{ij}(t) \right|^{2} 
\delta \left( \omega - \left(\varepsilon_{j}-\varepsilon_{i} \right) \right) \,  .
\ee
Note that $\omega$ is an energy throughout this paper. For a full round trip of the (dissipationless) trajectory, $V$ vanishes both in the initial and final (again unperturbed) state, and we may integrate by parts to obtain an expression for the spectrum of electron-hole pairs,
\bea
P_{\mathrm{ex}} (\omega) = \sum_{ij} \left| \frac{\lambda_{ij}}{\varepsilon_{j}-\varepsilon_{i}} \right|^{2} 
\delta \left( \omega - \left(\varepsilon_{j}-\varepsilon_{i} \right) \right)
\label{Pex}
\eea
with 
\be
\lambda_{ij} = \int_{-\tau}^{\tau} dt \, \left\langle \varepsilon_{j} \left| \frac{\partial V}{\partial t} \right| \varepsilon_{i} \right\rangle \exp(i(\varepsilon_{j}-\varepsilon_{i}) t /\hbar).
\label{lambda}
\ee
%
%
Extending our theory to finite electronic temperature by introducing Fermi occupation factors $f(\varepsilon)$, one has to consider the possibility of transitions where the metal electrons loose energy by going from occupied states above $\varepsilon_F$ to unoccupied states below $\varepsilon_F$. 
Thus, we need to define $P_{\rm{ex}} (\omega) = 0$ for $\omega < 0$. Following ref.~\onlinecite{HePe84}, eq.~(\ref{Pex}) needs to be generalized to 
\bea
\label{excitationspectra_elho}
P_{\rm{ex}} (\omega) &= &\sum_{ij} \left| \frac{\lambda_{ij}}{\varepsilon_{j}-\varepsilon_{i}} \right|^{2} 
\left(
f(\varepsilon_{i}) - f(\varepsilon_{j})
\right)
\nonumber\\
&& \times \delta \left( \omega - \left(\varepsilon_{j}-\varepsilon_{i} \right) \right)
\theta(\omega).
\eea
Interpreting defect electrons below the Fermi energy $\varepsilon_F$ as holes, first-order perturbation theory gives explicit expressions for the electron and hole spectra 
\bea
\label{excitationspectra_el_ho}
P_{\rm{ex, el}} (\omega) & = & \sum_{ij} \left| \frac{\lambda_{ij}}{\varepsilon_{j}-\varepsilon_{i}} \right|^{2} 
\left(
f(\varepsilon_{i}) - f(\varepsilon_{j})
\right)
\nn
&& \times
\delta \left( \omega - \left(\varepsilon_{j}-\varepsilon_{F} \right) \right)
\theta(\omega) \nonumber\\
P_{\rm{ex, ho}} (\omega) &=& \sum_{ij} \left| \frac{\lambda_{ij}}{\varepsilon_{j}-\varepsilon_{i}} \right|^{2} 
\left(
f(\varepsilon_{i}) - f(\varepsilon_{j})
\right)
\nn
&&\times
\delta \left( \omega - \left(\varepsilon_{i}-\varepsilon_{F} \right) \right)
\theta(-\omega).
\eea
The total energy loss of the trajectory is defined by 
\be
\Delta E = \int_{0}^{\infty} d \omega \, \omega \, P_{\rm{ex}} (\omega).
\label{Eloss}
\ee
In the following section, we will discuss the spectra for a half round trip in the adsorption well rather than a full one, in order to compare to previous results in the literature. For a dissipationless trajectory, the ways 'in' and 'out' are equivalent by time inversion symmetry, and within our approach we can take the spectrum of the half round trip just being equal to one half of the spectrum for the full round trip. 
Also, we note that time-reversal symmetry implies $\lambda_{ij}=0$ as $|\varepsilon_i - \varepsilon_j| \to 0$. 
But even without this assumption, the expression eq.~(\ref{lambda}) remains mathematically well-defined for a half round trip, as the matrix element in the integrand vanishes at the turning point of the trajectory. 
For evaluating $\lambda_{ij}$, we evaluate the integrand at discrete timesteps, and use
\be
\lambda_{ij} = \int_{-\tau}^{\tau} dt \, \left\langle \varepsilon_{j} \left| \frac{\Delta V(Q)}{\Delta Q} \right| \varepsilon_{i} \right\rangle \frac{dQ}{dt} \exp(i(\varepsilon_{j}-\varepsilon_{i}) t /\hbar).
\label{lambdaQ}
\ee
The DFT calculations are carried out at finite electronic temperature $T$, and the same temperature is used in the Fermi functions in eq.~(\ref{excitationspectra_elho}) and (\ref{excitationspectra_el_ho}). 
In order to plot the calculated spectra, it is necessary to replace the $\delta$-functions in eq.s~(\ref{excitationspectra_elho}) and (\ref{excitationspectra_el_ho}) by a function of finite width. 
In practice, we replace the $\delta$-function by $-\partial f / \partial \varepsilon$. The same electronic temperature $T$ as in the DFT calculations is used for the broadening. Matrix elements with $|\varepsilon_i - \varepsilon_j| < k_{\mathrm{B}} T$ have to be handled with care. Although the limit $|\varepsilon_i - \varepsilon_j| \to 0$ is mathematically well-defined, the numerical results are not reliable due to rounding errors. In the plots, and also in the calculations, this energy region is therefore omitted.
The total energy loss is not affected by this omission, since the contribution of such small-energy transitions to $\Delta E$ is negligible. 
(For a full round trip, $P_{\mathrm{ex}}(0)=0$). 

We note that the approach presented here can be extended beyond first-order perturbation theory. This is possible if the electron-hole pairs excited by the adsorbing particle can be treated as independent bosons. In this case, the electron-hole spectrum (but not the separate spectra for electrons and holes) can be calculated analytically. The method is applicable to weak 
and swift perturbations.\cite{MuRa71}
An extension to strong perturbations slowly varying in time has been suggested as well, where the wavefunctions used to calculate the matrix element in eq.~(\ref{lambdaQ}) are replaced by the ground-state wavefunctions calculated at each $Q$-value.\cite{ScGu81,ScGu82} For the system under study, we find that first-order perturbation theory is sufficient (see Section IV).
The theory presented here has some similarities
with, but also important differences to the theory of electronic friction worked out earlier.\cite{HePe84}
To see this, one assumes the separability of the matrix elements appearing in eq.~(\ref{lambda}) (see ref.~\onlinecite{TrBi03} for a discussion of this point), and takes the quasi-static limit. This limit implies that the excitation frequencies of interest are sufficiently small, such that the states $| \varepsilon_i \rangle$ and $\langle \varepsilon_j |$ are essentially indistinguishable from states at the Fermi surface, $| \varepsilon_F \rangle$. 
The resulting theory is formally identical to the theory of electronic friction, with the important difference that this theory uses wavefunctions changing adiabatically with the position of the adsorbate, while we use wavefunctions of the bare substrate to calculate the matrix elements.
We remind the reader that the theory of electronic friction starts from an expression for the imaginary part of dynamic self-energy $\Lambda(Q, \omega)$,
\bea
\mathrm{Im} \Lambda(Q, \omega)  &=& - \frac{2 \pi}{m} \sum_{i,j} \left| \left\langle \varepsilon_{j} \left| \frac{dV(Q)}{dQ} \right| \varepsilon_{i} \right\rangle \right|^2 
\nn
&& \times
\left(
f(\varepsilon_{i}) - f(\varepsilon_{j})
\right)
 \delta(\omega- (\varepsilon_{j}-\varepsilon_{i})) ,
\eea
see ref.~\onlinecite{HePe84}.
Note that this definition is somewhat analogous (although not equivalent) to the definitions in eq.~(\ref{lambda}) and~(\ref{excitationspectra_elho}). Whereas the present definition of $P_{\mathrm{ex}}$ uses the squared modulus of a time integral, the definition of Im$\Lambda$ starts from the square moduli of matrix elements. Hence a friction coefficient $\eta(Q)$ can be 
defined for any point $Q(t)$ along the trajectory by considering the limit $\lim_{\omega \to 0} m \Lambda(Q,\omega)/\omega $ and replacing the states $| \varepsilon_i \rangle$ and $\langle \varepsilon_j |$ by the states at the Fermi energy.
Taking into account the proper normalization of these states on the energy shell one can show that this limit exists, i.e., the friction coefficient is mathematically well-defined.
A similar reasoning employing the quasi-static limit can be used to establish that the total energy loss, eq.~(\ref{Eloss}), is a well-defined quantity also in our theory, both for a full and a half round trip. The limit of $P_{\mathrm{ex}}(\omega)$ for $\omega \to 0$ can be shown to exist, as the expression can be decomposed into the two factors,
$|\lambda_{ij}|^2 / |\varepsilon_j - \varepsilon_i |$ and $(f(\varepsilon_j) - f(\varepsilon_i))/ |\varepsilon_j - \varepsilon_i | $, 
which both have a finite limit for $| \varepsilon_i - \varepsilon_j| \to 0$. For the first factor, the proof is analogous to the derivation of the finiteness $\eta$ in electronic friction theory. The second factor is finite for any non-zero electronic temperature $T$. When considering the total energy loss, one may even send $T \to 0$; the combined limiting process $\omega \to 0; T \to 0$ for the integrand $\omega P_{\mathrm{ex}}(\omega)$ remains well-defined for any fixed ratio $\omega/(k_{\mathrm{B}}T)$.

In extension over electronic friction, the present theory allows 
to us address the excitations of a single electron-hole pair at {\em any finite} energy, even further away from the Fermi level. 
Compared to adsorbate vibrational damping, chemisorption is accompanied by stronger variations of the effective electronic potential, and hence components of higher frequency, also much higher than the vibrational frequency in the adsorption well, may become excited non-resonantly. 
The present theory is applicable under the following conditions: 
The energy loss $\Delta E$ during one round trip must be a small fraction of the total energy of the adsorbing particle, measured from the bottom of the adsorption well, to justify the 'unperturbed trajectory' approximation. Moreover, the particle is required to move sufficiently rapidly to justify the use of static (rather than adiabatic) electronic wavefunctions in evaluating the matrix elements. This means that the excitation frequency of interest, $\nu = | \varepsilon_j - \varepsilon_i | / h$, should fulfill the condition 
$\nu \gg (2 \pi \tau)^{-1}$, where $\tau$ is the duration of the impact (in scattering) or of half a round trip (in adsorption). The latter condition is fulfilled for electronic excitations detected by a Schottky diode ($h \nu > 0.5$~eV, i.e. $h \nu_{S} = 0.5$~eV) even for light particles such as thermal hydrogen atoms, where $\tau\sim20$ fs. 

Hence, the present theory seems to be more appropriate for the 
high-energy part of the excitation spectrum, detectable as
chemicurrents, than the electronic friction formalism. 
Moreover, the present, extended theory has some technical advantage concerning the so-called spin transition: 
In most practical implementations of DFT, the electronic correlation effects between the electrons at the adsorbing particle, e.g. hydrogen, and the metal electrons are described in an approximate way by solutions that break spin symmetry, by assigning a spin moment to the free hydrogen atom. During adsorption, spin symmetry is restored. In such ground-state DFT calculations, the spin transition has a formal analogy to a first-order phase transition, 
and thus leads to a $\sqrt{Q}$-type dependence of the effective potential near the transition point on the reaction path. In the theory of electronic friction, this singularity results in a divergence of $\Lambda(Q, \omega)$, rendering the theory inapplicable.\cite{TrBi03}
In the present theory, the $\lambda_{ij}$, and hence $P_{\mathrm{ex}}(\omega)$, remain well-defined, since the integrand in eq.~(\ref{lambdaQ}) diverges at most like $Q^{-1/2}$, and hence remains integrable. 
Phrased in other words, 
the abrupt change of the spin polarization in a ground-state DFT calculation and the quasi-static limit are incompatible.
The present theory circumvents this difficulty. 
However, we mention that in a time-dependent treatment of spin, both in TDDFT~\cite{LiPe06b} or in the mean-field approximation to the Newns-Anderson model,\cite{MiBi07} the spin polarization changes smoothly, and this difficulty doesn't arise anyway.

\section{Calculations}
The above concepts are illustrated by calculations for an H atom impinging on the on-top site on an Al(111) surface.
We chose this system because we can compare to recent results where the full dynamics has been treated within TDDFT.\cite{LiPe06b}
Here, we perform static DFT calculations within the generalized gradient spin density approximation (GGSDA). 
We employ the PBE-GGA exchange-correlation functional~\cite{PeBu96} together with norm-conserving pseudopotentials. 
We model the system by an Al(111) slab of $12$ layers with a $2 \sqrt{3} \times 2 \sqrt{3}$ unit cell. The slab is relaxed with respect to the interlayer distances, while we keep the lowest $3$ layers unrelaxed. Electronic wavefunctions are calculated using the {\em pwscf}-code~\cite{pwscf} for $35$ positions of an H atom above the surface on the on-top position of one of the substrate aluminium atoms, and for the substrate slab alone. The cut-off energy for the plane-wave expansion of the wavefunctions is chosen to be $21$~Ry, and a $4 \times 4 \times 1$ Monkhorst-Pack {\bf k}-point mesh is used for the Brillouin zone integration. The cut-off energy was tested by calculating the chemisorption energy of H for the on-top position on Al(111), once using the cutoff of $21$ Ry, yielding $1.83$ eV, and once using a cutoff of $40$ Ry, yielding a chemisorption energy of $1.86$ eV. The difference of $30$ meV is acceptable in our opinion, when compared to the much larger calculation times for the higher cutoff. TDDFT calculations did not even reach this cutoff if a reasonable calculation time is a necessity, but the results from static calculations compare well to our numbers ($1.90$ vs. $1.87$ eV for $40$ and $20$ Ry cutoff, respectively).\cite{LiPe06b}
A smearing of the occupation factors of $1$ mRy using a simple Fermi-Dirac broadening scheme is used, which corresponds roughly to the experimental temperatures.\cite{KrNu07a} %
Using the aformentioned parameters we obtained a lattice constant of $4.061$~{\AA} for bulk Al, while the experimental value is $4.050$~{\AA}. 

\begin{figure}[t]
\centering

\includegraphics[width=8.6cm]{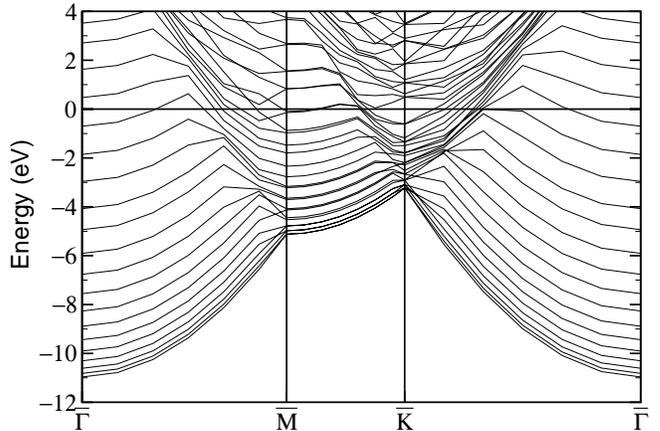}

\caption{\label{bandstructure} Kohn-Sham band structure of an Al(111) slab consisting of $12$ layers in a $(1 \times 1)$ unit cell calculated within PBE-GGA. A Monkhorst-Pack {\bf k}-point mesh of $16 \times 16 \times 1$ points has been used to calculate the charge density and the self-consistent potential. The Fermi energy is set to 0.
}
\end{figure}

Fig.~\ref{bandstructure} shows the Kohn-Sham band structure of an Al(111) slab with a $(1 \times 1)$ unit cell resulting from our calculations. Comparison with the literature~\cite{JaZh02} shows good agreement. The lowest state at the $\overline{\Gamma}$-point is about $11$~eV below, the band bottom at the $\overline{M}$-point about $5$~eV, and at the $\overline{K}$-point about $3$~eV below the Fermi energy. 
%

\begin{figure}[t]
\centering

\includegraphics[width=8.6cm]{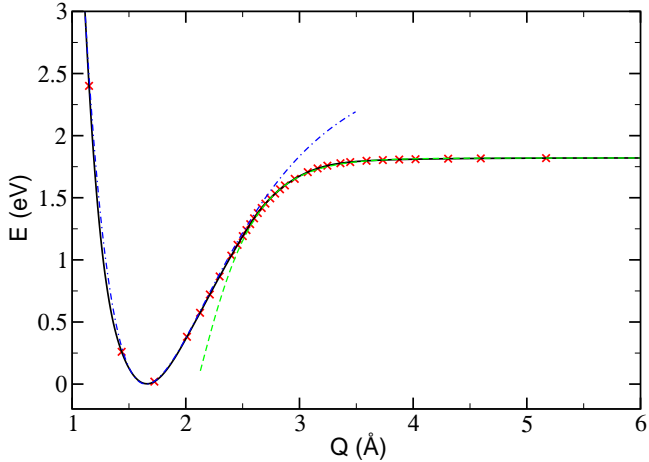}

\caption{\label{potential}(Color online) Chemisorption potential of a H atom above the on-top position of an Al(111) surface. 
The spin polarization in the ground-state DFT calculations vanishes 
at about $2.6$~{\AA}. The dashed lines are two Morse potentials fitted to either branch of the PES.}
\end{figure}

\begin{figure}[t]
\centering

\includegraphics[width=8.6cm]{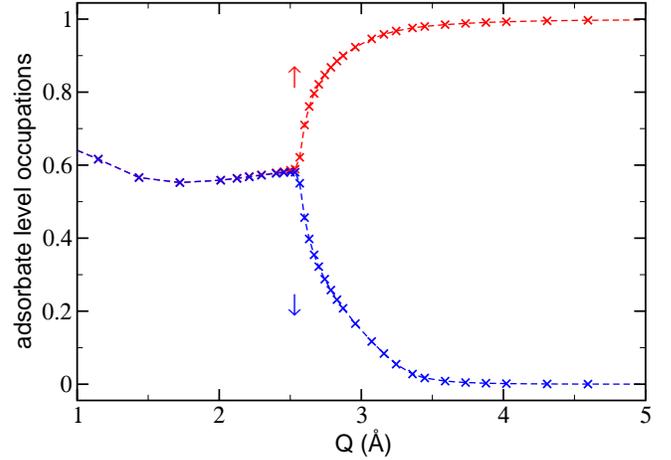}

\caption{\label{spintransition}(Color online) Ground-state occupation of the spin-up and spin-down orbital of an approaching H atom vs. its distance from an Al(111) surface. Far away from the surface, the electronic charge at the H atom, corresponding to one electron, is fully spin-polarized. Both spin-up and spin-down orbitals become equally populated after the spin transition point. The spin polarisation shows a $\sqrt{Q}$-behavior close to the transition point.}
\end{figure}

Fig.~\ref{potential} shows the chemisorption potential of hydrogen above the on-top site of Al(111). The ground-state potential energy surface (PES) consists of two branches that intersect at a distance of about $2.6$~{\AA} above the surface. For larger distances, the DFT ground state is spin-polarized, while the electronic system is non-polarized when the H atom comes closer. As seen in fig.~\ref{spintransition}, the electronic spin polarization at the H atom vanishes below this point.

Fig.~\ref{spintransition} shows the result of a population analysis for the atomic $1s$ orbital of the H atom using the L\"owdin approach implemented in {\it pwscf}. 
The DFT ground-state wavefunctions of either spin for each position along the trajectory were projected onto the H $1s$ orbital. 
Below the spin transition point at about $2.6$~{\AA} both the spin-up and the the spin-down orbital shows a population of about 0.6~electrons. 
Further away from the surface, the H atom is spin-polarized, i.e., the spin-up orbital gets fully populated while the spin-down orbital gets depleted. Right above the transition point the spin polarization shows a sharp, square-root-like increase.

The classical trajectory for vertical impact of the H-atom onto the on-top site of Al(111) is calculated from Newton's equations for an initial kinetic energy of $60$~meV, and starts 3.16 {\AA} above the surface. Thereby, the small effect of electronic friction on the trajectory is neglected. The total time needed for one half round trip is 20 fs.

The numerical evaluation of eq.~(\ref{lambdaQ}) proceeds as follows: 
The derivative $\partial V/ \partial Q$ is calculated numerically, using the first derivative of a third-order spline interpolation. The potential used is the effective potential from the ground-state DFT calculations. The matrix elements in eq.~(\ref{lambdaQ}) are computed in two steps: First we perform a summation in real space to calculate $\Delta V | \varepsilon_i \rangle$, then we make a Fast Fourier Transform to wave-vector space and multiply by $\langle \varepsilon_j |$. The wavefunctions are those of the initial state at $Q(-\infty)$, but omitting adsorbate wavefunctions. Likewise, the energies in the exponential function in eq.~(\ref{lambdaQ}) are the Kohn-Sham energies of the initial state. The time integration is then done using a spline interpolation of the matrix elements. For the technical implementation the state indices $i$ and $j$ incorporate band index, {\bf k}-point and spin. Only transitions within the same {\bf k}-point and spin channel are considered. However, the restriction regarding the {\bf k}-points can be lifted.\cite{TrGr01} A weight factor for the summation over {\bf k}-points also has to be introduced.
%
\section{Results}
The main result of the theory presented here are the spectra of the electrons and holes excited by a hydrogen atom impinging with thermal kinetic energy. From the part of the spectra exceeding the Schottky barrier, we directly derive the predicted yield of the measured chemicurrent. In order to study the isotope effect systematically, we not only carry out calculations for H and D, but also for fictitious atoms of mass $0.1m_H$ and $0.01m_H$. The spectra are shown for a half round trip in the adsorption potential (until the particle has reached the turning point on the repulsive branch of the PES), to be able to compare with earlier results.\cite{BiMi08}
We note that in the experiments the particles perform many round trips before they have dissipated the chemisorption energy and come to rest. 
Hence the average number of electron-hole pairs per incident hydrogen atom in an actual experiment will be significantly higher than the values displayed in the figures. 

\begin{figure}[t]
\centering

\includegraphics[width=8.6cm]{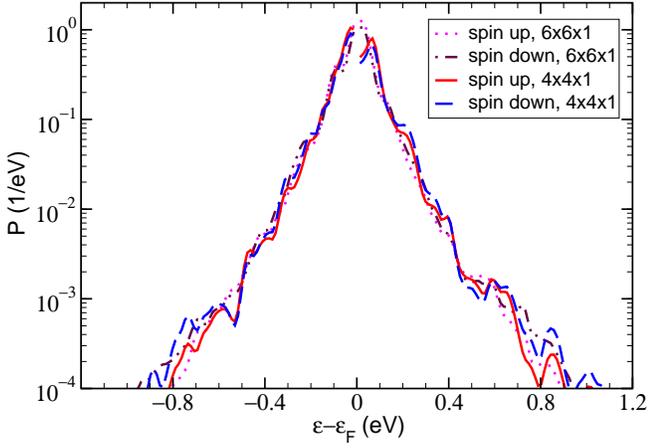}

\caption{\label{spectrum_H}(Color online) Spin-resolved excitation spectra of electrons (positive energies) and holes (negative energies) for an impinging H atom with an initial kinetic energy of $60$~meV on the on-top position of an Al(111) surface after one half round trip in the chemisorption well. 
Values close to the Fermi energy ($ |\varepsilon - \varepsilon_F | < 0.013$~eV) are not plotted because of numerical inaccuracies in this energy range. }
\end{figure}

\begin{figure}[t]
\centering

\includegraphics[width=8.6cm]{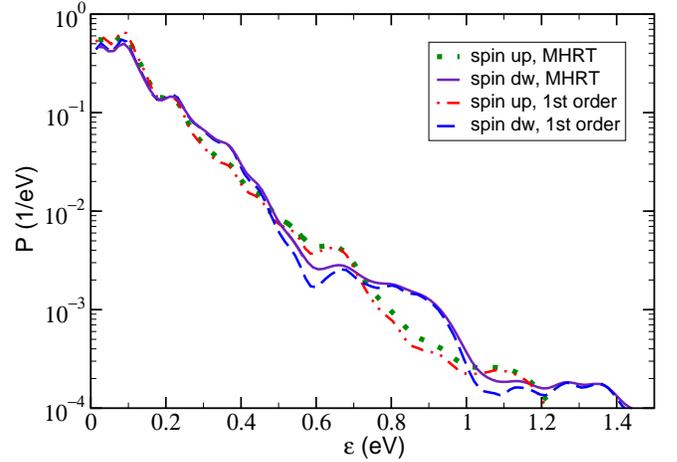}
\caption{\label{Spectrum_elho_H}(Color online) Spin-resolved excitation spectrum of electron-hole pairs for an impinging H atom with an initial kinetic energy of $60$~meV on the on-top position of an Al(111) surface after one half round trip in the chemisorption well. 
The curves calculated in first-order perturbation theory are in good agreement with those obtained by a resummation of the perturbation expansion following M{\"u}ller-Hartmann {\em et al.} (MHRT).
Values close to zero energy ($ \varepsilon < 0.013$~eV) are not plotted (and omitted in the calculations) because of numerical inaccuracies in this energy range.
}
\end{figure}

Fig.~\ref{spectrum_H} shows the spectra of electrons (positive energies) and holes (negative energies) for a half round trip of an H atom. We notice that the spectra of electrons and holes are quite similar. Moreover, differences in the spectra of spin-up electrons (same spin polarization as the impinging H atom) and spin-down electrons (opposite spin polarization as the impinging H atom) are only marginally different when plotted on a logarithmic scale. The same holds for the holes of either spin. 
The full and dashed lines in fig.~\ref{spectrum_H} indicate the results of convergence tests. While a metal has a continuous spectrum of excitations, the slab used in the DFT calculations has a large but finite number of states, due to the finite thickness and the number of {\bf k}-points used to sample the Brillouin zone. For convergence tests, we increased this sampling from a $4 \times 4 \times 1$ {\bf k}-point mesh to a $6 \times 6 \times 1$ mesh. 
(This refers to the Kohn-Sham wavefunctions $| \varepsilon_i \rangle $ used to calculate the matrix elements in eq.~(\ref{lambdaQ}); the potential is assumed to be converged already for the smaller {\bf k}-point set.)
Fig.~\ref{spectrum_H} shows that the spectra are almost identical, albeit the $6 \times 6 \times 1$ calculations gives somewhat smoother curves.
The energy loss $\Delta E$ is 23~meV for the smaller {\bf k}-point set and 20~meV for the larger one. 
The error appears to be tolerable, and for the following calculations, the smaller set was used.
Note that the energy loss is just a small fraction, about 1\%, of the chemisorption energy, see fig. \ref{potential}. Hence, the use of an elastic trajectory in calculation of the matrix elements appears to be justified.
Fig.~\ref{Spectrum_elho_H} shows the spectrum of the electron-hole pairs. 
Here we compare the results of first-order perturbation theory, eq.~(\ref{excitationspectra_elho}), to the treatment by M{\"u}ller-Hartmann and co-workers\cite{MuRa71} that sums up all orders in the perturbation expansion. We find that the two approaches yield very similar spectra when plotted on a logarithmic scale. This demonstrates that first-order perturbation theory is sufficient for the high-lying excitations with small excitation probability, which are of interest to the calculation of chemicurrents. Another remark is of interest here: 
If only the combined electron-hole spectrum were known, one would be tempted to assume that electrons and holes are distributed symmetrically around the Fermi energy. Comparing to fig.~\ref{spectrum_H}, we find that this assumption is reasonable in the present case (but not for the stronger excitations by fictitious atoms lighter than H, see below). 

\begin{figure}[t]
\centering

\includegraphics[width=8.6cm]{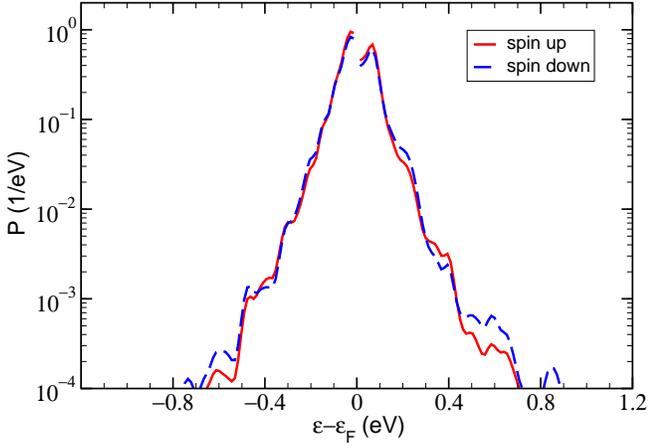}

\caption{\label{Spectrum_D}(Color online) Spin-resolved excitation spectra of electrons and holes for D with an initial kinetic energy of $60$~meV after one half round trip in the chemisorption well. The spectrum is qualitatively similar to the spectrum for H, shown in fig.~\ref{spectrum_H}, but the decay is much faster.}
\end{figure}

\begin{figure}[t]
\centering

\includegraphics[width=8.6cm]{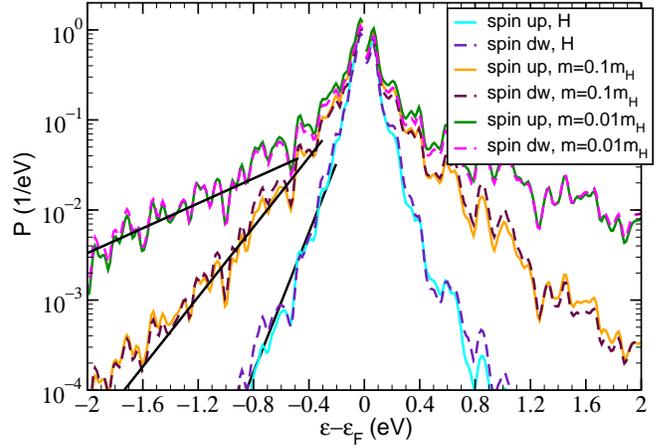}

\caption{\label{spectra_isotopes}(Color online) Spin-resolved excitation spectra of electrons and holes for different impinging isotopes of H with an initial kinetic energy of $60$~meV after one half round trip in the chemisorption well. Spectra for three different masses with $m=m_{H}$, $m=0.1m_{H}$ and $m=0.01m_{H}$ are shown, where lower masses give spectra with a larger amplitude. Decaying exponentials with $A \exp (- \varepsilon / k_{B} T_{\mathrm{eff}})$ are fitted to the hole part of the spectra. There is some freedom in choosing the energy window for the fit, and hence the slopes are associated with large error bars. The lines shown in the plot correspond to $T_{\mathrm{eff}}= 1300$~K, $2600$~K, and $7300$~K, respectively.}
\end{figure}

Next, we investigate the isotope effect. 
The electron and hole spectra for deuterium are shown in fig.~\ref{Spectrum_D}. They are qualitatively similar to those of H, but the excitation probability decays more strongly with increasing excitation energy of the charge carriers. 
Inspired by the work of Lindenblatt and Pehlke,\cite{LiPe06c} 
we also used fictitious isotopes of smaller mass, 
$0.1 m_{H}$ and $0.01 m_{H}$, in order to get stronger excitations that facilitate the analysis. The spectra are collected in fig.~\ref{spectra_isotopes}. 
Similar to previous work,\cite{LiPe06c} we fit an 'effective temperature' $T_{\mathrm{eff}}$ to the high-energy tails of the spectra, using a two-parameter fit $P(\varepsilon) = A \exp (- \varepsilon / k_{B} T_{\mathrm{eff}} )$. For the hole spectra, the fitted regions are shown in fig.~\ref{spectra_isotopes}. The temperatures obtained from the fit are given in table~\ref{temperature}. 

\begin{table}[h]
\renewcommand\arraystretch{1.2}
\begin{tabular}{|l|cc|cc|}
\hline
Isotope & \multicolumn{2}{|c|}{Perturbation theory}  & \multicolumn{2}{|c|}{TDDFT~\cite{LiPe06c}} \\
& $T_{\mathrm{eff}}^{e}$ [K] & $T_{\mathrm{eff}}^{h}$ [K] & $T_{\mathrm{eff}}^{e}$ [K] & $T_{\mathrm{eff}}^{h}$ [K] \\\hline\hline
Deuterium & $1100 \frac{-300}{+400}$ & $1000 \frac{-200}{+300}$ &  &  \\
Hydrogen & $1400 \frac{-300}{+700}$ & $1300 \frac{-300}{+700}$ & $900\frac{-100}{+100}$ & $1100\frac{-100}{+100}$ \\
$H_{\rm{Exp}}$ &  & $1680$~\cite{KrNu07a,KrNu07b} & & \\
$H: m=0.1 m_{H}$ & $3000 \frac{-500}{+1400}$ & $2600 \frac{-500}{+1400}$ & $2700\frac{-200}{+200}$ & $2300\frac{-200}{+200}$ \\
$H: m=0.01m_{H}$ & $8900 \frac{-3800}{+300}$ & $7300 \frac{-2900}{+200}$ & $9300\frac{-200}{+200}$ & $7800\frac{-2200}{+200}$ \\
\hline%
\end{tabular}
\caption{\label{temperature}Slope parameter $T_{\mathrm{eff}}$ of the electron and hole spectra for different isotope masses. The middle column contains the fits to our data, and the right column contains fits to data from TDDFT calculations.\cite{LiPe06c} The value $H_{\rm{Exp}}$ is the experimental value for H adsorption on Ag. The error bars for the fits are quite large. The reason is that the fits can be made over a wide energy area, starting and ending on different points, and thus are rather arbitrary in a certain range. 
}
\renewcommand\arraystretch{1.0}
\end{table}

\begin{figure}[t]
\centering

\includegraphics[width=8.6cm]{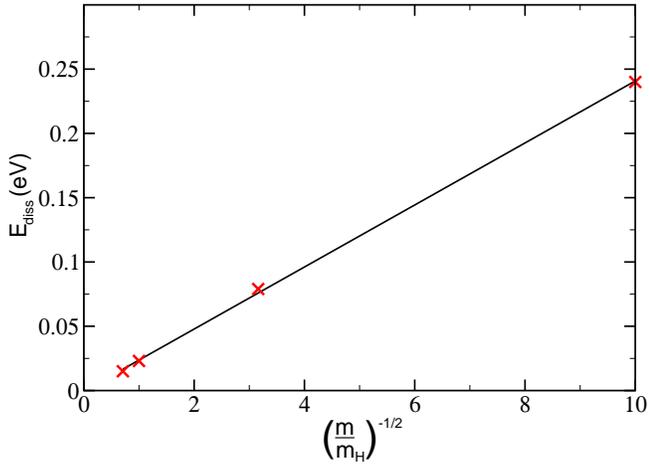}

\caption{\label{eloss_isotopes}(Color online) Total energy loss vs. the mass scaling parameter $1/ \sqrt{m/m_{H}}$. $m$ denotes the mass of the adsorbing atom. The line is a linear fit to the data. 
}
\end{figure}

The quite large error bars were estimated by varying the fitting window. They indicate that the distributions are not truly exponential. A similar uncertainty of the fitted temperatures was also noticed in previous work.\cite{LiPe06c,BiMi08} We note that the temperatures fitted for the excited electrons are systematically higher than for the holes. This is particularly obvious for the light isotopes, but also holds for H and D, despite the nearly symmetric appearance of the spectra in fig.~\ref{spectrum_H}. 
Within the error bars, the temperatures obtained from perturbation theory are in reasonable agreement both with experiment and with the TDDFT calculations.
Although the effective temperatures are helpful for identifying trends, 
an analysis of the isotope effect on this basis is not conclusive due to the large error bars, and it is safer to study the isotope effect in the total energy loss, and in the integrated probability of excited carriers above the Schottky barrier. Fig.~\ref{eloss_isotopes} shows that the linear relationship between the total energy loss and the mass scaling parameter $(m/m_H)^{-1/2}$.
This can be rationalized in view of electronic friction theory. In this theory, the friction coefficient depends only on the electronic properties of the system, and hence would be the same for all isotopes of the same chemical element. The frictional force is thus  proportional to the velocity of the impinging particle. By plotting the energy loss as a function of $(m/m_H)^{-1/2}$, one would therefore expect all data points to fall on a straight line. This also holds in the present, more elaborate theory.
The integrated probabilities $N_e$ and $N_h$ of excited carriers above the Schottky barrier are shown in table~\ref{detected_particles}, where a barrier height of 0.48~eV has been assumed, typical of experimental Schottky barriers for Ag on Si.\cite{KrNu07b,NiBe99}

\begin{figure}[t]
\centering

\includegraphics[width=8.6cm]{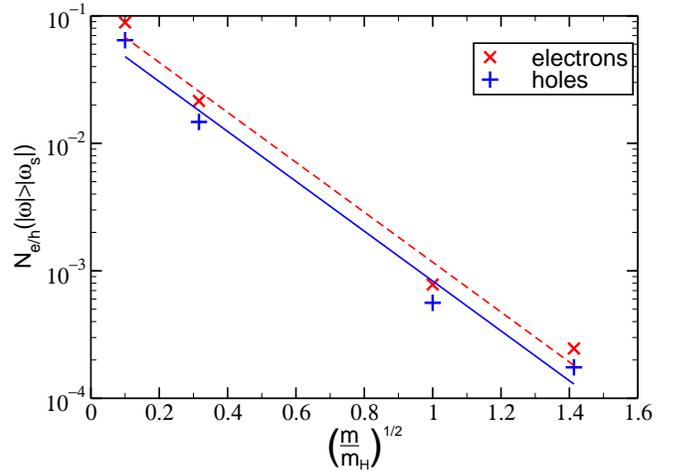}
\caption{\label{yield_vs_m}(Color online) Yield of the electron (red, upper curve) and hole (blue, lower curve) excitation spectra plotted versus the mass-scaling parameter $\sqrt{m/m_{H}}$. $m$ denotes the mass of the adsorbing atom. The lines are linear fits to the data. }
\end{figure}

\begin{table}[h]
\begin{tabular}{|l|c|c|} 
\hline 
Isotope & Electrons & Holes \\\hline\hline
Deuterium & $0.000246$ & $0.000175$ \\ \hline
Hydrogen & $0.000781$ & $0.000561$ \\ 
$N/N_{D}$ & $3.17$ & $3.21$ \\ 
$N/N_{D,\rm{Experiment}}$ & & $3.7 \pm 0.7$~\cite{KrNu07a,KrNu07b} \\ \hline
$H: m=0.1 m_{H}$ & $0.021502$ & $0.014713$ \\ 
$N/N_{H}$ & $27.53$ & $26.23$ \\ \hline
$H: m=0.01m_{H}$ & $0.088775$ & $0.064376$ \\ 
$N/N_{H}$ & $113.67$ & $114.75$ \\ \hline
\end{tabular}
\caption{\label{detected_particles}Average number of electrons and holes created per impinging H isotope with an initial kinetic energy of $60$~meV, per half round trip. 
}
\end{table}
In electronic friction theory, the excitation probabilities are assumed to be Poissonian distributions~\cite{TrBi03}. All moments of such a distribution a solely determined by its first moment. It follows that all excitation spectra (of different isotopes) for a given chemical species impinging on a given substrate are described by a one-parameter family of curves. Then, a single parameter, e.g. the energy loss $\Delta E$, is sufficient to characterize the spectra completely. Here, we test if such a simple description of the spectra also applies to the results of our theory. The scaling properties of $\Delta E$ have already been established above.
If one supposes a proportionality of the effective temperatures $T_{\mathrm{eff}}$ of the high-energy tail and the total energy loss $\Delta E$, one would expect that a logarithmic plot of $N_{e}$ or $N_{h}$ shows a linear dependence on $(m/m_H)^{1/2}$. 
Figure~\ref{yield_vs_m} demonstrates that this is a good approximation for masses between $2m_H$ and $0.01m_H$, i.e., it is fair to approximate the energy spectra by a single-parameter family of curves. 
The ratio of the integrated probabilities $N_{e/h}$ for H and D atoms is about 3.2.
(3.17 if calculated for the excited electrons, 3.21 for the holes, see table~\ref{detected_particles}). 
Experimentally, the hole excitation probabilities are more accessible, 
since Schottky barriers on p-type silicon can be fabricated with a high quality of the metal-semiconductor interface.
The experimental ratio of the yields for H and D for Ag/Si Schottky diodes is $3.7 \pm 0.7$,\cite{KrNu07a,KrNu07b} which agrees well with our results. 
%

%

\section{Discussion}
At present, the possibilities to test our results against experiment are rather limited. Quantitative comparison is possible so far only for the isotope effect: Here, our results for Al are in good agreement with the measurements using Ag/Si-Schottky diodes~\cite{KrNu07a,KrNu07b} and MIM sensors~\cite{MiHa06}.
Some experimental information is also available about the slopes of the excitation spectra from measurements with Schottky diodes made from the same material but with slightly different {\it I-V}-characteristics. 
The temperatures estimated from these measurements are in reasonable agreement with the slopes of our calculated spectra. 
However, we are quite sceptical whether a single temperature would still be a good description if one had better experimental access to the spectrum. 

A more detailed comparison of our results is possible with previous TDDFT calculations for the same system, H on Al(111). In non-spinpolarized calculations, a total energy loss of 30 meV and 40 meV was found after a half and a full round trip, respectively.\cite{LiPe06c} This is in reasonable agreement with the values of 23 meV and 46 meV calculated in our approach. 
In addition, the linear dependence of the energy loss on the mass scaling parameter (cf. fig.~\ref{eloss_isotopes}), and the effective temperatures of the spectra (cf. table~\ref{temperature}) obtained in both theories are consistent.
However, more recent spin-polarized TDDFT calculations yielded energy losses of about 100 meV and 185 meV for the half and full round trip, respectively.\cite{LiPe06b} These findings from the TDDFT calculations suggest that the spin transition is quite important for the energy loss.
While our theory in principle includes the spin transition, it seems to underestimate its role in the energy dissipation. Preliminary calculations where we use a smoothened and delayed spin transition (similar to the findings of ref.~\onlinecite{LiPe06b}) instead of an abrupt one show a higher energy loss in perturbation theory, too.
Moreover, in the TDDFT calculations, it was assumed for technical reasons that the total spin polarization (integrated over both the hydrogen atom and the substrate) is conserved, while in our ground-state DFT calculations the total spin polarization changes from about one electron for the hydrogen far away from the surface to zero after hydrogen adsorption.
This methodological difference might be the reason why Lindenblatt and Pehlke find a shift (in energy) between the spin-up and spin-down spectra~\cite{LiPe06b} after a half round trip, while our spectra are almost identical for both spins. However, this discrepancy does not arise after a full round trip, where spin-dependent features are absent in both methods. 
%
%
We may also compare our results to earlier work on H/Cu(111) using the electronic friction formalism.\cite{TrGr02,TrGr01,TrBi03}
This theory assumes that the energy loss proceeds by multiple excitation of very small energy quanta, with the excitation energy of a single quantum ultimately tending to zero, while the total energy loss remains finite in this limit.
Under this assumption, the energy loss due to friction can be described by an instantaneous friction force in each moment of time, and the materials properties of the substrate only need to be known in vicinity of the Fermi energy in order to calculate the friction coefficient.
This is different from our approach, where we calculate the probability to deposit a finite amount of energy to the substrate within a single excitation. 
Both approaches are complementary: While our method focusses on the spectral properties of the excitations, 
including electronic band structure details above or below the Fermi energy, our formulae are applicable only to a half or full round trip of the particle, i.e. at turning points of the trajectory. 
The electronic friction formalism, on the other hand, allows 
one to break down the energy loss (at least within the limitations implied in this formalism) to every point along the particle's trajectory.
In order to obtain the excitation spectra from the friction approach, additional assumptions need to be made, 
such as the separability of matrix elements, and the statistical independence of multiple excitations.
The spectra (for H/Cu(111)) obtained after making these approximations are essentially exponential distributions. 
Our results from first-order perturbation theory are qualitatively similar, but cannot be fit by a single exponential. This implies that the picture of multiple excitations made up from energy quanta close to zero may be an oversimplification.
We note that a connection between both approaches can be made on the basis of the overall energy loss per round trip they predict.
From the more detailed spectra obtained from our approach, it is in principle possible to determine {\it a posteriori} an effective friction coefficient that would lead to the same amount of energy loss as obtained by integrating over our excitation spectra. 
This effective friction coefficient could then be used to re-run the trajectory to obtain new spectra, and the procedure could be iterated until consistency between the energy loss of the trajectory and the loss implied by the spectra has been reached.
 
Recently, non-adiabatic effects in adsorption have been studied in the framework of a time-dependent Newns-Anderson model treated in mean-field approximation.\cite{MiBi05,MiBi07,MiBi07E,BiMi08,MiBi08}
While this model is originally motivated by the surface chemistry of transition metals and noble metals, it may still make sense to compare the results to our calculations for H on Al(111). The model requires a fit to the data supplied by DFT calculations, i.e. the projected density of states onto the H orbitals. In ref. \onlinecite{BiMi08}, Mizielinski {\it et al.} obtained spectra for H/Al(111) that show clear differences between electrons and holes, and between both spin orientations. However, the results are rather sensitive to the model parameters. One can observe that the parametrization of the values needed for the model calculations does not fit perfectly, especially far away from the surface where ionization and affinity level of H are about equally far below or above the Fermi energy of Al.\cite{BiMi08} For this case, Mizielinski {\it et al.} predict electron and hole spectra that are approximately equal, and which have no important spin characteristics, cf. ref. \onlinecite{MiBi08} for metals other than Aluminium. We think it can be justified to argue that H on Al(111) also corresponds in the language of this model to such a system, so that one would expect electron and hole spectra that are approximately equal. This is found in our calculations, too.
However, their model also predicts that electron and hole spectra may be strongly different for other combinations of substrate and adsorbate. 
In order to probe the different regions of the parameter space of the Newns-Anderson model discussed in ref.~\onlinecite{MiBi07}, i.e., strongly electronegative or electropositive adsorbates, additional DFT studies for such material systems would be desirable.

\section{Conclusions}
In this work, we implemented first-order time-dependent perturbation theory to calculate excitation spectra of electrons and holes in a metal substrate due to non-adiabatic effects in adsorption, starting from the ground-state electronic potential and wavefunctions obtained from density functional theory. 
The calculated results for adsorption of hydrogen on aluminum are consistent with the experimental data obtained for hydrogen on noble metals. In particular, the present theory reproduces the observed isotope effect. 
In contrast to the electronic friction formalism, our approach includes excitations over finite energy differences, and thus takes the effects of single-particle band structure into account in the excitation spectra.
Our results are in agreement with results from the (computationally much more demanding) TDDFT method, with the notable exception of spin effects. On the other hand, our method using perturbation theory is much faster and affordable for a wider range of systems, as it requires only ground-state DFT calculations. These calculations are widely used in the scientific community, e.g. in the calculation of potential energy surfaces within the Born-Oppenheimer approximation.
The method presented here can be implemented in existing (static) DFT codes as a post-processing tool, and is applicable in cases where the full dynamics of the many-particle system remains close to the Born-Oppenheimer surface. 

\section*{Acknowledgments} 
The authors thank D.M. Bird, E. Pehlke, M. Mizielinksi, and M. Lindenblatt for fruitful discussions. We acknowledge granted computer time from the John-von-Neumann center for scientific computing (NIC) in J\"ulich, Germany. Financial support is granted by SFB 616 of the DFG, ''Energy dissipation at surfaces''.

%

\end{document}